# Magnetoelastic instabilities in kagome antiferromagnet $Mn_{3-x}Ga$

Linxuan Song,[1,2] Feng Zhou,[1,2] Guilin Lu,[1] Liang Yao,[1,2] Xuekui Xi,[2] Yong-Chang Lau,[2,3*] Youguo Shi and Wenhong Wang[1*]

[1]School of Electronics and Information Engineering, Tiangong University, Tianjin 300387, China

[2]Institute of Physics, Chinese Academy of Sciences, Beijing 100190, China

[3]University of Chinese Academy of Sciences, Beijing 100049, China

*Author to whom correspondence should be addressed: yongchang.lau@iphy.ac.cn, wenhongwang@tiangong.edu.cn

**Abstract**

We present a systematic study of the structural, magnetic, and transport properties of hexagonal $Mn_{3-x}Ga$ alloys, revealing a series of composition-controlled emergent phenomena. By tuning the Mn concentration, we uncover distinct lattice responses, including a zero thermal expansion–like volume compensation behavior in Mn-poor compositions and a magnetoelastic-driven, field-assisted structural phase transition in Mn-rich samples. These lattice instabilities are accompanied by correlated magnetic and transport anomalies, including metamagnetic transitions, negative magnetoresistance, and anomalous Hall sign reversal. First-principles calculations demonstrate that the Hall sign reversal originates from crystal-symmetry breaking rather than magnetic reorientation alone. Our results establish composition as the key control parameter governing magnetoelastic coupling in $Mn_{3-x}Ga$, providing a unified framework to tailor structural, magnetic, and topological transport properties in kagome antiferromagnets and reconcile previously disparate experimental observations.

Kagome non-collinear antiferromagnetic materials, in particular Mn$_3$X (X = Ga, Ge, Sn) have attracted considerable attention in spintronics due to their distinctive spin structures and properties [1-3]. The triangular spin alignment in the Mn$_3$X alloys results in nearly zero net magnetization [4, 5]. Moreover, the Mn-atom-based kagome lattice, coupled with spin-orbit interaction from the magnetic structure, induces a topological electronic band structure that imparts remarkable anomalous transport properties [6-10] which can be manipulated by strain [11-14], current [15] and composition [16, 17]. These unique characteristics make Mn$_3$X alloys a compelling focus for both experimental and theoretical research in antiferromagnetic spintronics, with promising applications in ultra-fast information processing [18-20].

Off-stoichiometric Mn$_3$Ga is a newly confirmed material exhibiting the highest anomalous Hall conductivity among Mn$_3$X alloys, attributed to Fermi-level tuning by excess Ga atoms [21]. This discovery follows nearly a decade after large anomalous transport properties were observed in Mn$_3$Sn [6] and Mn$_3$Ge [7], despite challenges in sample preparation of single crystal due to the complex Mn-Ga phase diagram [22] and the existence of multiple Mn$_3$Ga structures [4, 23-27]. Transport signature of the Weyl fermions in off-stoichiometric Mn$_3$Ga has also been observed by planar Hall effect accompanied by the negative magnetoresistance [28]. Notably, hexagonal Mn$_3$X alloys are all stabilized exclusively in off-stoichiometric forms: Mn$_3$Ga stabilizes with Mn deficiency [26, 29], while Mn$_3$Sn and Mn$_3$Ge tend to form with Mn excess [2, 16],. Recent studies showed that element doping can stabilize single-phase Mn$_3$Sn [30-33] and Mn$_3$Ge [34-37] leading to tunable magnetic and transport properties, with varying spin reorientation with the doping content, revealing the effectiveness of chemical engineering in tailoring nontrivial spin textures and the anomalous Hall effect (AHE) in Mn$_3$X alloys.

Hexagonal Mn$_3$Ga, however, maintains a pure phase over a wide composition range at room temperature without doping [26], offering a simple pathway to manipulate its magnetic and transport properties. Nonetheless, the composition dependence of its magnetic and transport properties remains unclear. Unlike the Mn$_3$Sn and doped Mn$_3$Ge [35, 37] which only exhibit magnetic phase transition at low temperature [38, 39], the crystal structure of hexagonal $Mn_3Ga$ shows a slightly distortion at low temperatures [29]. Interestingly, the nature of this distortion varies across reported studies. The distortion was initially observed in XRD spectra as peak splitting around 80 K [29] and was subsequently linked to a transition in the AHE [40]. However, in our previous work, although a slight distortion was also observed, neither was a peak splitting detected in the XRD spectra, nor was a significant change observed in the AHE [41]. In this letter, we report a systematic study of the structure, magnetism and the transport properties of hexagonal off-stoichiometric $Mn_{3-x}Ga$ with varying compositions (x=0.65-0.2). The different temperature-dependent structural features observed in previous studies of the hexagonal $Mn_3Ga$ will also be discussed.

Polycrystalline hexagonal $Mn_{3-x}Ga$ ribbons with nominal compositions x=0.2-0.65 were prepared by arc-melting and subsequent melt-spinning under the high-purity argon atmosphere. Then, the sample was annealed at 600 °C in an evacuated quartz tube for one week and quenched in water to obtain the pure hexagonal phase. The annealed $Mn_{3-x}Ga$ (x=0.2-0.5) ribbons were subsequently ball-milled and re-annealed to obtain powder samples for temperature-dependent structural characterization. The crystal structure was characterized using X-ray diffraction (XRD) using Cu Kα radiation equipped with a variable-temperature stage. The magnetic properties were measured on a magnetic property measuring system (MPMS, Quantum Design) with high-temperature capability. Transport properties were measured on a physical property measuring system (PPMS, Quantum Design). The first principles Calculations were conducted using density functional theory (DFT) implemented in the Vienna ab-initio simulation package [42] (VASP) code. The exchange correlation functional is the Generalized-Gradient-Approximation [43] (GGA) of the Perdew-Burke-Ernzerhof [44] (PBE) functional.

The hexagonal off-stoichiometric $Mn_{3-x}Ga$ can be synthesized with Mn compositions ranging from approximately 70 to 73.6 at.%, corresponding to the composition range $Mn_{2.35}Ga$–$Mn_{2.8}Ga$ (x= 0.2-0.65). Beyond this range, the tetragonal phase emerges in samples with higher Ga content, while the cubic phase forms in samples with an excess of Mn [26]. The XRD pattern and Rietveld refinements results of hexagonal $Mn_{3-x}Ga$ (x=0.2, 0.65) ribbons at room temperature are shown in Fig.1 (a) and (c). The XRD pattern and Rietveld refinements results of other $Mn_{3-x}Ga$ alloys are presented in Fig. S1 (See the Supplementary Materials). All diffraction peaks are indexed to be $D0_{19}$ type hexagonal structure with space group $P_{63}/mmc$ (194). In off-stoichiometric $Mn_{3-x}Ga$ alloy system, the Mn deficiency can either lead to vacancies, or being occupied by excess Ga atoms (Ga-rich), which can be distinguished based on the intensities of specific XRD peaks. Notably, the intensity of the (110) peak changes significantly, while the intensity of the (220) peak remains unchanged depending on the occupation of atoms (See the Supplementary Materials). The intensity ratio of (220), (110) peaks in the XRD patterns of hexagonal $Mn_{3-x}Ga$ alloys with varying compositions, derived from both experimental observations and calculations that account for excess Ga atoms occupying Mn vacancies and only Mn vacancies in off-stoichiometric $Mn_3Ga$ is shown in Fig. 1(b). The experimental intensity ratio closely matches the calculated ratio that accounts for Ga atoms occupying Mn vacancies, indicating that the off-stoichiometric $Mn_3Ga$ alloys are Ga-rich systems. Furthermore, as shown in Fig. 1(d), the lattice parameters decrease in general, with increasing Mn content. This behavior is likely due to Ga atoms, which have a larger atomic radius, occupying Mn sites.

The temperature dependence of magnetization (*M-T curves*) for hexagonal $Mn_{3-x}Ga$ alloys is shown in Fig 2(a). The *M-T curves* exhibit two inflections: The paramagnetic transition temperature ($T_N$) around 430 ~ 480 K which increases with increasing Mn content [21], and an additional magnetic transition temperature ($T_d$) observed at low temperature, where the magnetization increase steeply. The latter may correspond to a structural distortion, as reported in previous studies of hexagonal $Mn_3Ga$ alloys [41, 45]. Fig. 2(b) shows the Mn composition dependence of $T_d$ for hexagonal $Mn_{3-x}Ga$ alloys. As Mn content decreases, $T_d$ decreases and the magnetic transition become more gradual. A comparison of the M–T curves with those reported in the literature suggests that the discrepancy among the former reports [29, 40, 41] may originate from differences in sample composition. It is worth noting that the $Mn_{2.8}Ga$ (x=0.2) sample exhibits additional magnetic features distinct from the other compositions: the *M–T curve* shows an extra magnetic transition around 160 K, which suggest that the additional changes in the magnetic state may be involved.

The field dependence of the magnetization (*M-H curves*) for two typical alloys $Mn_{2.35}Ga$ (x=0.65) and $Mn_{2.8}Ga$ (x=0.2) at various temperatures are shown in Figs. 2(c) and 2(d). At room temperature, the *M-H curves* exhibit a linear response, indicating antiferromagnetism with a small, hysteretic net magnetization at zero field. A metamagnetic transition occurs near the $T_d$, reflecting a transition from antiferromagnetism to ferromagnetism. These variations are consistent with the behavior observed in $Mn_{2.48}Ga$ [41]. *M-H curves* for other $Mn_{3-x}Ga$ alloys are presented in Fig. S2, as Mn content decreases, the metamagnetic transition fades gradually (see the Supplementary Materials). Specifically, the *M-H curves* go through three stages: Firstly, above the $T_d$, all samples maintain the antiferromagnetism, showing the linear response to the external filed. Secondly, as the temperature approaches $T_d$, the antiferromagntic spin structure is unstable due to the lattice contraction at low temperature and the spin tends to rotate along with the higher magnetic field more easily, triggering the metamagnetic transition. Finally, below $T_d$ the antiferromagnetic spin configuration is completely suppressed, resulting in ferromagnetic-like behavior. In low-Mn content samples, the transition only reaches the second stage, exhibiting trends towards metamagnetic transition but not fully achieving it, possibly due to the weaker lattice distortion.

To clarify the structural evolution of $Mn_{3-x}Ga$ around the characteristic temperature, temperature-dependent X-ray diffraction measurements were performed on $Mn_{3-x}Ga$ powder samples. The temperature-dependent XRD patterns of the $Mn_{3-x}Ga$ are shown is Fig.S3 (See Supplementary Materials). Among the investigated compositions, only $Mn_{2.8}Ga$ (x=0.2) exhibits a clear splitting of diffraction peaks at low temperatures around 160 K, far below $T_d$ of about 260 K, indicating a structural phase transition from the hexagonal phase to the monoclinic phase, a subgroup of the orthorhombic structure in terms of space-group symmetry [46]. This structural transition coincides with the additional magnetic anomalies observed in $Mn_{2.8}Ga$, suggesting a common origin. In contrast, low-Mn content $Mn_{3-x}Ga$ samples show no diffraction-peak splitting near $T_d$, excluding a symmetry-breaking structural phase transition associated with the metamagnetic behavior. Instead, only nonmonotonic peak shifts are observed upon cooling, indicating that the magnetic anomalies near $T_d$ originate from a lattice-related response. Since the low-temperature structural phase transition of Mn-rich $Mn_3Ga$ has been well documented in previous studies, the present work focuses on the evolution of the crystal lattice near $T_d$ in $Mn_{3-x}Ga$, where lattice distortions occur without an accompanying structural phase transition, providing a suitable platform to investigate the coupling between lattice and magnetic degrees of freedom in the noncollinear antiferromagnetic state.

The temperature dependent lattice constants and the unit cell volume of hexagonal $Mn_{3-x}Ga$ (x=0.2-0.5) are shown in Fig.3. More detailed Rietveld refinements results are shown in Fig.S4 (See Supplementary Materials). The $T_d$ is marked as the red symbols. Remarkably, in the low-Mn content region, the lattice parameters and unit-cell volume exhibit an almost temperature-independent behavior over a finite temperature window around $T_d$, indicative of a near-zero thermal expansion [47, 48], which is a newly observation in the hexagonal non collinear antiferromagnets $Mn_3X$ (X=Ga, Ge, Sn) system. The zero thermal expansion emerges only in proximity to the magnetic transition temperature $T_d$ and coincides with the metamagnetic instability of the noncollinear antiferromagnetic state, pointing to a magnetic-transition-driven lattice response rather than a conventional volume compensation mechanism. The onset of this anomalous lattice behavior occurs at temperatures close to $T_d$, demonstrating a direct correlation between magnetic ordering and the suppression of thermal expansion. In contrast, Mn-richer compositions show a smoother, more conventional thermal evolution of lattice parameters across $T_d$, with no pronounced zero-expansion regime. These results establish zero thermal expansion as an intrinsic lattice response in low-Mn content $Mn_{3-x}Ga$ near $T_d$, highlighting an unusually strong magnetoelastic coupling in the noncollinear antiferromagnetic state. In addition, a closer comparison of the unit-cell volume reveals that, except for the low-Mn compositions exhibiting near-zero thermal expansion, the values of V at $T_d$ are remarkably similar across different compositions. This observation suggests the existence of a characteristic lattice state associated with the magnetic transition. Despite the compositional dependence of the absolute lattice parameters, the onset of lattice distortion near $T_d$ occurs when lattice contraction reaches a comparable threshold, implying that the magnetic transition is governed by a common criterion. This behavior indicates that the evolution of magnetic interactions in $Mn_{3-x}Ga$ follows a unified trend, predominantly controlled by the Mn–Mn interatomic distances reduced by lattice contraction. In this picture, the lattice distortion near $T_d$ acts to partially accommodate the exchange-driven reconfiguration of magnetic interactions, while the low-temperature symmetry-lowering structural phase transition observed in Mn-rich compositions represents an extreme case in which the magnetoelastic coupling can no longer be relieved by continuous lattice distortion and instead requires a breaking of crystallographic symmetry.

By correlating the structural and magnetic properties of hexagonal $Mn_{3-x}Ga$ alloys, it becomes evident that the magnetic behavior of off-stoichiometric $Mn_3Ga$ is governed by a strong coupling between lattice contraction and exchange interactions. In $Mn_{3-x}Ga$, partial substitution of Mn by larger Ga atoms leads to an overall lattice expansion with decreasing Mn content. Upon cooling, the progressive lattice contraction reduces the Mn–Mn interatomic distances, thereby continuously tuning the frustrated exchange interactions among Mn atoms in the kagome plane. The characteristic temperature $T_d$ corresponds to a critical lattice state, marked by a nearly composition-independent unit-cell volume, at which the balance between lattice elasticity and magnetic exchange interactions is altered. At this point, the noncollinear antiferromagnetic configuration becomes unstable, giving rise to the metamagnetic transition observed in the M–H curves. In the vicinity of $T_d$, the exchange-driven instability can be accommodated by a continuous lattice distortion without crystallographic symmetry breaking. However, for Mn-rich compositions close to $Mn_3Ga$, further lattice contraction drives the system beyond the accommodation limit of continuous distortion. Once a critical threshold is exceeded, magnetoelastic coupling can no longer be sustained within the original symmetry, leading to a symmetry-lowering structural phase transition at low temperatures, as observed in $Mn_{2.8}Ga$.

The transport properties of $Mn_{3-x}Ga$ alloys also exhibit clear composition-dependent behavior. $T_d$ is also evident in the temperature dependence of longitudinal resistivity (*R-T* curves), as shown in Fig.S5 (see the Supplementary Materials). The MR of $Mn_{3-x}Ga$ alloys at 300K and 10K are shown in Fig. 4(a), and the other MR of the $Mn_{3-x}Ga$ alloys at other temperatures is shown in Fig. S6(see the Supplementary Materials). All the samples show positive MR value at room temperature corresponding to the stable, noncollinear antiferromagnetic state, while the MR behavior at 10K becomes strongly composition dependent and can be classified into three distinct regimes. For the low-Mn content samples (x=0.65, 0.6), the MR remains positive over the entire field range, indicating that the antiferromagnetic spin configuration is largely preserved even under strong magnetic fields. In contrast, the Mn-rich sample (x=0.2) shows a purely negative MR, consistent with a ferromagnetic-like state stabilized by a symmetry-lowering structural transition at low temperatures. For other intermediate compositions, the MR exhibits a nonmonotonic field dependence, with a weak positive MR at low fields followed by a crossover to negative MR at higher fields. The relatively small magnitude of the negative MR suggests that, although these compositions undergo a metamagnetic transition, the associated magnetic reconfiguration is moderate and only evolves toward a ferromagnetic-like state under sufficiently strong magnetic fields. This behavior is fully consistent with the structural analysis discussed above, where intermediate compositions experience lattice distortion without crystallographic symmetry breaking, while a pronounced ferromagnetic response and negative MR emerge only after the lattice symmetry is broken. Notably, the metamagnetic transitions observed in the *M–H curves* are also reflected in the corresponding MR curves as field-induced changes in magnetoresistance. The MR values for various compositions under a 5 T magnetic field across different temperatures are summarized in Fig. 4(b). Except for $Mn_{2.35}Ga$ and $Mn_{2.4}Ga$, the MR exhibits a peak near $T_d$ and turns negative at lower temperatures, consistent with the magnetoelastic-driven magnetic evolution discussed above.

Fig.4 (c)-(e) show the Hall resistivity of polycrystalline $Mn_{3-x}Ga$ (x=0.65, 0.4, 0.2) measured at different temperatures, highlighting a systematic evolution from room temperature to low temperature that correlates closely with the magnetic and structural behaviors discussed above. At room temperature, all compositions exhibit a conventional anomalous Hall of the noncollinear antiferromagnetic state, where the suppression of $\rho_H$ at high fields reflects field-induced destabilization of the spin texture, leading to a reduction of the intrinsic anomalous Hall contribution. This trend persists upon cooling toward the distortion temperature $T_d$. For low-Mn content samples, the relatively weak lattice distortion leads only to a further reduction of the anomalous Hall signal at high fields without a sign change. In contrast, with the Mn composition increasing, samples display a more pronounced evolution of the Hall response, leading to a field-driven sign reversal of $\rho_H$ at low temperatures, indicating a substantial reconstruction of the magnetic structure. For compositions close to $Mn_3Ga$, the low-temperature structural phase transition results in a complete reversal of the Hall signal, consistent with a symmetry-broken magnetic ground state. Notably, for each composition, the zero-field anomalous Hall resistivity reaches a maximum near $T_d$, suggesting an additional contribution possibly associated with complex spin textures, arising a topological Hall component.

Notably, the Hall effects reported in Liu et al.'s work [40] and our previous study [41] align with those observed for $Mn_{2.8}Ga$ (x=0.2) and $Mn_{2.4}Ga$ (x=0.6), respectively. In other words, at low temperatures, the lattice distortions should contribute to the Hall resistivity through multiple mechanisms: the topological Hall effect induced by complex spin textures [34, 49, 50], AHE originating from the distorted hexagonal phase, and the superposition of anisotropic AHE from different crystal orientations. Further studies are need to quantitatively separate these contributions through detailed transport measurements using single crystal samples. However, the sign reversal of the anomalous Hall effect is analyzed qualitatively based on first-principles calculations for different magnetic and structural configurations.

According to neutron diffraction results, the monoclinic phase transition only induces an in-plane rotation of a local moment, leading to an enhanced net in-plane magnetic moment compared with the hexagonal phase [46]. Motivated by this picture, we calculated the anomalous Hall conductivity (AHC) of $Mn_3Ga$ for three configurations: the noncollinear antiferromagnetic state in the hexagonal structure, the in-plane local-moment-rotated state within hexagonal symmetry, and the monoclinic structure. As shown in Fig. 5, the AHC vanishes along the magnetic hard-axis for all configurations. Within the hexagonal symmetry, the AHC along high-symmetry directions in the easy-magnetization plane exhibits the same sign for both magnetic configurations, excluding magnetic reorientation without symmetry lowering as the origin of the Hall sign reversal. In contrast, the monoclinic structure shows opposite AHC signs along different in-plane directions, evidencing a symmetry-breaking–induced reconstruction of the Hall response. For polycrystalline relevance, we further performed orientational averaging of the AHC for each configuration (Fig. 5d). Considering the upward Fermi-level shift caused by Ga substitution, we focus on the energy window from the Fermi level up to +0.051 eV, corresponding to $Mn_{2.43}Ga$ [21]. Within this range, a sign reversal appears exclusively in the monoclinic phase, demonstrating that the Hall sign change is intrinsically driven by crystal-symmetry breaking rather than magnetic rotation alone.

In summary, we present a systematic investigation of the structural, magnetic, and transport properties of hexagonal $Mn_{3-x}Ga$ alloys, revealing a series of composition-controlled properties. By tuning the Mn content, we uncover distinct lattice responses, including a zero thermal expansion behavior in the low-Mn content region and a magnetoelastic-driven, field-assisted structural phase transition in Mn-rich compositions. These composition-dependent lattice instabilities are accompanied by correlated magnetic and transport anomalies, such as metamagnetic transitions, negative magnetoresistance, and anomalous Hall sign reversal. Our results demonstrate that the diverse and previously reported behaviors in $Mn_3Ga$-based systems originate from composition-induced magnetoelastic coupling, providing a unified framework that reconciles apparent discrepancies in the literature.

*Acknowledgements.*

This work was supported by the National Natural Science Foundation of China (Grants Nos. 12504156, 12274321, 12274438, and 12361141823), National Key R&D program of China (No. 2022YFA1402600), Beijing Natural Science Foundation (Grants Nos. Z230006 and IS25044). A portion of this work was carried out at the Synergetic Extreme Condition User Facility (SECUF).

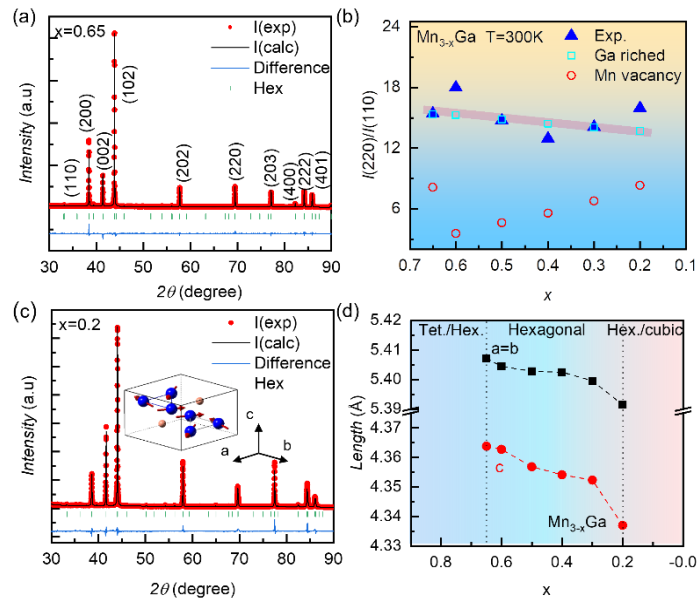

Figure.1 (a) The XRD pattern and Rietveld refinements results of hexagonal $Mn_{3-x}Ga$ ribbons alloys. (b) Peak intensity ratio of (220)/(110) peaks in XRD patterns of hexagonal $Mn_xGa$ alloys with varying compositions, derived from experimental observations and calculations that account for excess Ga atoms occupying Mn vacancies and Mn vacancies in off-stoichiometric $Mn_3Ga$. (c) The phase diagram of hexagonal $Mn_{3-x}Ga$ ribbons. The inset shows the lattice parameters of pure hexagonal $Mn_{3-x}Ga$ alloys.

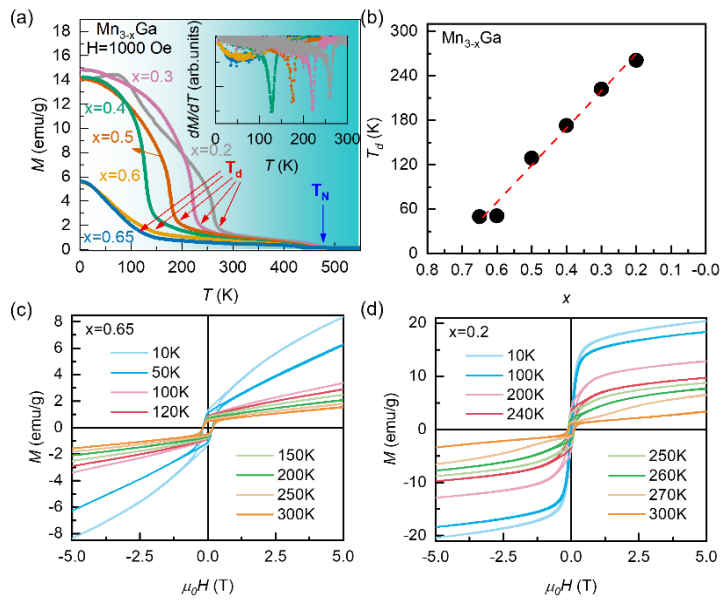

Figure.2 (a) Temperature dependence of magnetization (*M-T curves*) of the hexagonal $Mn_{3-x}Ga$ alloys. The inset shows the derivation of the M-T curves. (b) The crystal distortion temperature

($T_d$) of the hexagonal Mn$_{3-x}$Ga. Magnetic field dependence of the magnetization (*M-H curves*) of the Mn$_{2.35}$Ga (c) and Mn$_{2.8}$Ga (d).

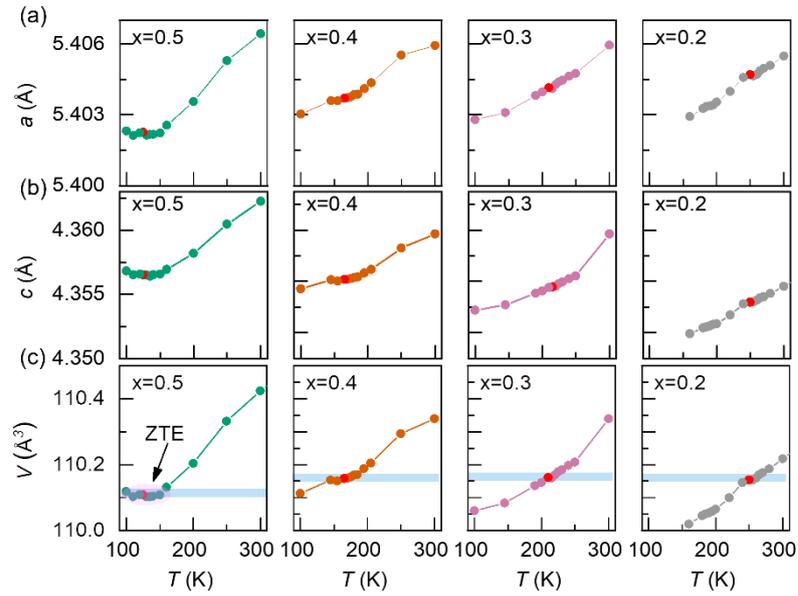

Figure.3 The Rietveld refinements results of hexagonal Mn$_{3-x}$Ga powders. (a) The temperature dependence of the lattice constant a, (b) lattice constant c of Mn$_{3-x}$Ga and (c) volume of the unit cell. (The $T_d$ of Mn$_{3-x}$Ga is shown as red marker.)

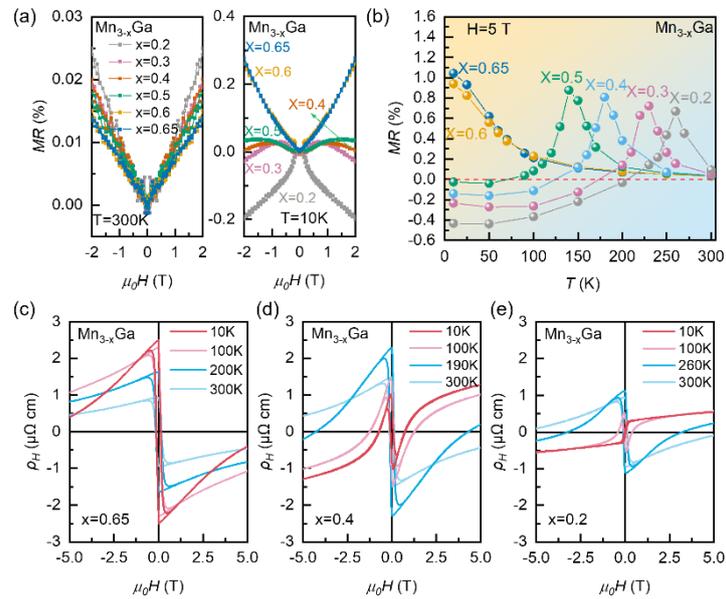

Figure. 4 (a) Magnetic field dependences of the magnetoresistance (MR) of Mn$_{3-x}$Ga at 10 K and 300 K. (d) The MR of the Mn$_{3-x}$Ga at various temperature at 5T. Magnetic field dependences of the Hall resistivity of Mn$_{3-x}$Ga, (c) x=0.65, (d) x=0.65 and (e) x=0.2.

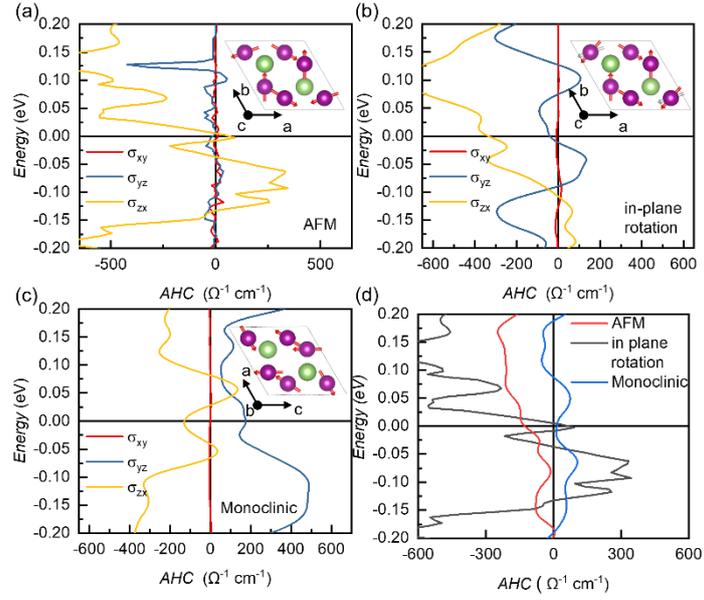

Fig 5 Intrinsic anomalous Hall conductivity (AHC) of $Mn_3Ga$ in different structural or magnetic forms: the noncollinear antiferromagnetic configuration (a) as well as the in-plane local-moment-rotation state (b) in hexagonal structure and the monoclinic structure; (d) Orientationally averaged AHC for each configuration.